\documentstyle[prb,
aps,psfig]{revtex}
\begin{document}
\draft

\twocolumn

\title{Density functional study of the 
adsorption of K on the Ag(111) surface}
\author{K. Doll}
\address{Institut f\"ur Mathematische Physik, TU Braunschweig,
Mendelssohnstra{\ss}e 3, D-38106 Braunschweig}
\maketitle

\begin{abstract}
Full-potential gradient corrected density functional calculations
of the  adsorption of potassium on the Ag(111) surface have been 
performed. The considered structures are 
Ag(111)$(\sqrt 3 \times \sqrt 3)$R30$^\circ$-K and Ag(111)$(2\times 2)$-K.
For the lower coverage, fcc, hcp and bridge site; and
for the higher coverage all considered sites are practically degenerate.
 Substrate rumpling is most
important for the top adsorption site. The bond length is found to be
nearly identical for the two coverages, in agreement with recent experiments.
Results from Mulliken populations, bond lengths, core level shifts and 
work functions consistently indicate a small charge transfer from the
potassium atom to the substrate, which is
slightly larger for the lower coverage.
\end{abstract}

\pacs{ }

\narrowtext
\section{Introduction}

The study of adsorbates on metal surfaces has become a center of
interest in surface science because of the enormous importance of
catalysis for
industrial applications. These experimental studies can be complemented 
with simulations which 
are nowadays a powerful tool for the theoretical description of surface
structures and chemical reactions at surfaces, because of the
growth in computational power and due to the success of density
functional calculations. 

One prototype reaction
is the adsorption of alkali metals on metallic surfaces.
Although this is a relatively simple reaction, 
there have been several surprises
in the study of these systems. Initially, it was assumed that
the adatoms would occupy hollow sites, until
a first system was discovered where top site adsorption was
preferred \cite{Lindgren1983}. 
Meanwhile, further systems of alkali metals
with top site 
adsorption have been found --- typically heavier alkali atoms (from 
K on) on close packed surfaces (for reviews, see, e.g., references 
\onlinecite
{DiehlMcGrathlang,DiehlMcGrathkurz,Over,Zangwill,StampflScheffler,SchefflerStampfl}). 
One important finding
was that top site adsorption was accompanied by substrate rumpling and
thus it was assumed that substrates with a low bulk modulus would 
be prominent for top site adsorption.
It was therefore a little
surprise when in a LEED (low energy electron diffraction)
study of K on Ag(111) it was discovered that
hollow sites were occupied
\cite{Leathermanetal,Kaukasoinaetal}.

The type of hollow site occupied in the system K/Ag(111)
depends on the coverage: in the low coverage
structures ((2 $\times$ 2) and (3$\times$3)), the adatoms occupy
face centered cubic (fcc) hollow sites, in the 
$(\sqrt{3}\times\sqrt{3})$R30$^\circ$ structure 
the hexagonal close packed (hcp) hollow 
sites\cite{Leathermanetal,Kaukasoinaetal}. In addition, the
phase diagram of K, Rb and Cs adsorbed on Ag(111) was investigated
experimentally \cite{LeathermanDiehl1996}.

There have been several simulations of alkali metals on close-packed
metallic surfaces:
for example,
systems such as Na and K on the Al(111) surface
\cite{NeugebauerScheffler}, 
Na on the Cu(111) surface \cite{CarlssonHellsing}, K on the Pt(111)
 surface \cite{Moreetal} and K on Cu(111)
\cite{Doll2001KCu} have been studied. 
The system K/Ag(111) is therefore a very interesting system due to
the experimental findings and as an extension of the employed technology 
(section \ref{ParameterSection})
to substrates of the second row of the transition metals. 

It is therefore the aim of this article to summarize results from
 simulations on
the system K/Ag(111). Full-potential density-functional
calculations with a local Gaussian
basis set and with a gradient corrected functional were performed,
for two coverages (1/3 or 1/4 of a monolayer,
for a $(\sqrt{3} \times \sqrt{3})$R30$^\circ$ or (2 $\times$ 2) pattern).
The addressed questions are the  preferred adsorption 
sites,
the magnitude of the
energy splitting of different highly symmetric adsorption sites,
the geometry and the importance of surface rumpling,
the charge of the adsorbate, the work function
and the positions of the K core levels.

\section{Computational Parameters}

\label{ParameterSection}

A local basis set formalism was used where the basis functions are
Gaussian type orbitals centered at the atoms as implemented in the code
 CRYSTAL \cite{Manual}. For Ag, a relativistic
small-core pseudopotential (with 28
electrons in the core) in
combination with a $[4s3p2d]$ basis set was 
employed as in a previous study of Cl on Ag(111) \cite{DollHarrison2001}.
The $[5s4p1d]$ K all-electron basis from reference \onlinecite{Doll2001KCu} 
was chosen.
The exchange-correlation potential
was fitted with auxiliary basis sets as in previous
work \cite{Doll2001KCu,DollHarrison2001}.
The gradient corrected exchange and correlation functional of Perdew,
Burke and Ernzerhof (PBE) \cite{PBE} was used.

The adsorption was modeled by using slabs of 4 or 5 silver layers,
at the PBE-optimized bulk Ag
lattice constant\cite{DollHarrison2001} of $a_0=4.10$ \AA.
Potassium was adsorbed on one side of this slab. 
A supercell approach with a ($\sqrt{3} \times \sqrt{3}$)R30$^\circ$ or
(2$\times$2) structure as in the experiment was implemented. 
This slab was truly two dimensional and thus
{\em not} periodically repeated in the third dimension.
The vertical relaxation of the silver atoms was simulated
 in three different ways:
simulations were performed where only a uniform relaxation of the top
silver layer was possible, and simulations where a different vertical
relaxation of the silver atoms in the top layer was possible (i.e. 
substrate rumpling). 
Finally, an additional lateral displacement of the atoms in the top
silver layer, as observed experimentally
for K/Ni(111)\cite{Fisheretal}, was simulated.

\begin{figure}
\caption{The structures considered for K, 
adsorbed on the Ag(111) surface, at a coverage of one third of a monolayer,
$(\protect\sqrt 3 \times \protect\sqrt 3)$R30$^\circ$ unit cell. 
The silver atoms in the top layer 
are displayed by open circles. The considered potassium adsorption sites are
the top site above the silver atoms with number 1 (filled circles),
the threefold hollow 
sites above atoms 1,2,3 (fcc or hcp hollow, possibly with stacking fault,
circles with horizontal lines) 
or the bridge site above atoms 2 and 3
(circles with horizontal and vertical lines). Note that the threefold hollow
sites can not be distinguished in this figure. }
\label{geometryfigure}
\centerline
{\psfig
{figure=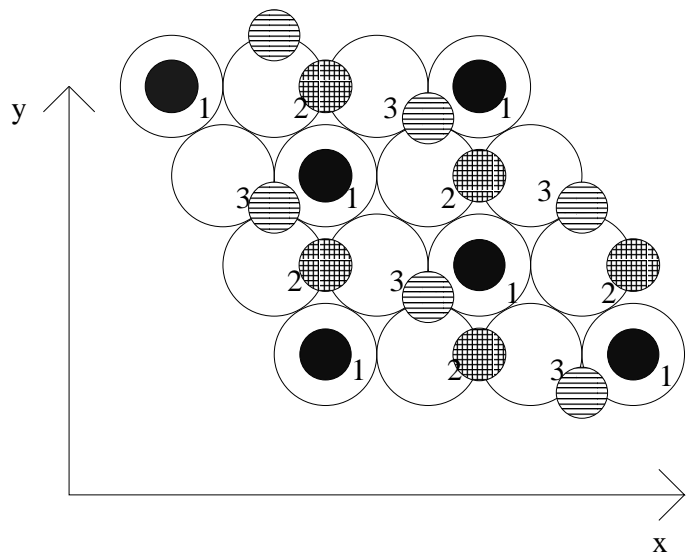,width=15cm,angle=270}}
\end{figure}

\begin{figure}
\caption{The structures considered for K, 
adsorbed on the Ag(111) surface, at a coverage of one fourth of a monolayer,
$(2\times 2)$ unit cell. The silver atoms in the top layer 
are displayed by open circles. The considered potassium adsorption sites are
the top site above the silver atoms with number 1 (filled circles),
the threefold hollow 
sites above atoms 2,3,4 (fcc or hcp hollow, possibly with stacking fault,
circles with horizontal lines) 
or the bridge site above atoms 3 and 4
(circles with horizontal and vertical lines). Note that the threefold
hollow sites can not be distinguished in this figure. }
\label{geometryfigure2}
\centerline
{\psfig
{figure=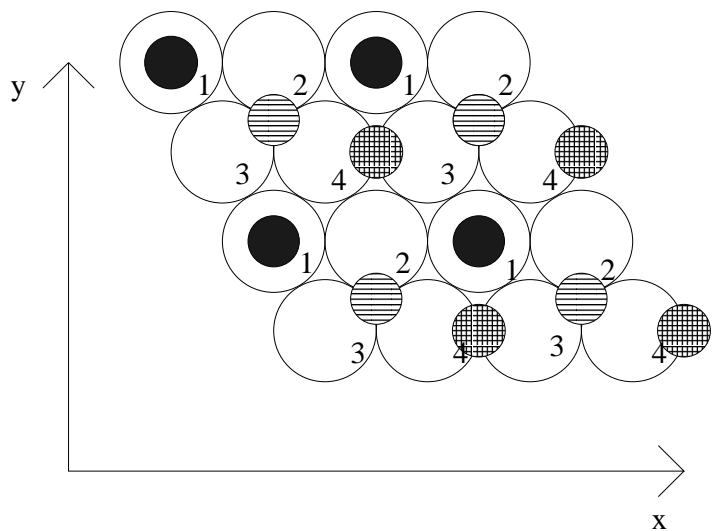,width=15cm,angle=270}}
\end{figure}

Four adsorption sites were considered (see figures
\ref{geometryfigure} and \ref{geometryfigure2}): 
the top adsorption site with K sitting vertically
above a silver atom in the top layer, the bridge site with K sitting 
above the middle of two silver atoms in the top layer, and two
different threefold hollow sites
where the potassium atoms are placed vertically above 
a silver atom in the second (third) silver layer (hcp and fcc hollow,
respectively).
In addition, the possibility of a stacking fault was investigated. 
In this structure, the outermost Ag layer and the K layer would occupy
hcp sites.
Such a stacking fault was observed \cite{Soares} for the system Ag(111)
 $(\sqrt{3} \times \sqrt{3})$R30$^\circ$-Sb. 

\begin{figure}
\caption{Definition of the geometrical parameters. All distances are
interlayer distances (definition see text). Note that in all the figures,
the atoms are drawn purely schematically and the
 size of the atoms does not scale with the atomic radii.}
\label{geometryfigure3}
\centerline
{\psfig
{figure=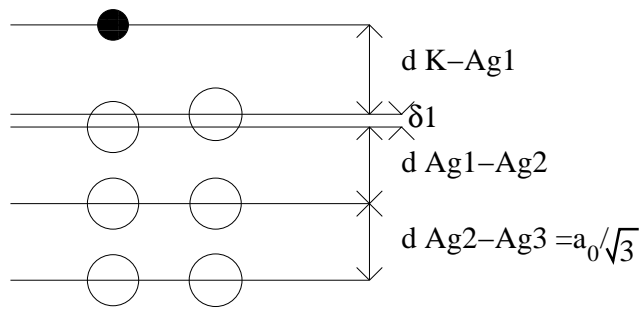,width=15cm,angle=270}}
\end{figure}

The optimized geometrical parameters
are defined as in figure \ref{geometryfigure3}. The variables are chosen
in such a way so that the notation is consistent with those from
 the experimental references 
\cite{Leathermanetal,Kaukasoinaetal}.
The bond  length is defined as
the distance from a potassium adatom to the nearest silver atom.
$d_{K-Ag1}$, $d_{Ag1-Ag2}$, $d_{Ag2-Ag3}$
and $\delta 1$ are various interlayer distances:
$d_{K-Ag1}$ is the vertical distance from the potassium layer to the
plane made of the atoms in the first silver layer which have moved towards
the potassium layer. $d_{Ag1-Ag2}$ is the distance between the plane made of
the atoms in the  first layer which have moved towards
the second Ag layer, and the second Ag layer. $d_{Ag2-Ag3}$, the distance
between second and third layer silver atoms, was kept fixed at the value
of $4.10/\sqrt{3}$ \AA \
corresponding to the distance of two layers in the bulk ($a_0/\sqrt{3}$).
Finally,
$\delta 1$ is the rumpling, i.e. the distance between the plane made of
those atoms in the first silver layer which have moved towards the K atoms
and the plane made of those atoms in the first silver layer which have moved
towards the second silver layer. 
Additionally, the possibility of a lateral displacement of the atoms
in the top Ag layer was taken into account (for a more detailed description
of the lateral displacement, see
the following section \ref{potassiumadsorptionsection}).

A net with 16 $\times$ 16 sampling points in the
surface Brillouin zone was used \cite{PackMonkhorst}.
During the optimization process, 
the Fermi function was smeared with a temperature of 0.03 $E_h$
(1$E_h$=27.2114 eV) and the total energy
extrapolated to zero temperature\cite{Gillan}. 
To ensure the stability of the results with respect
to this parameter, additional single-point
calculations with a smearing temperature
of 0.005 $E_h$ were performed (see also appendix). Also, the results for
properties such as Mulliken population, work function etc. were computed with
a smearing of 0.005 $E_h$.
For an extensive test of the various computational
parameters for metallic systems with
the CRYSTAL code, 
see also reference \onlinecite{KlausNicVic}.

The unit cells which 
are used in the simulations for this article are at the limit of what
is presently feasible with this method.
This local basis set approach is thus for metals probably more
expensive than alternative approaches relying on plane wave technologies.
However, it is interesting to have the possibility of comparing
various strategies which complement each other. 
One advantage of the method employed here is that
concepts such as computing Mulliken charges are easily implemented in
a code relying on Gaussian basis sets. Also, all-electron calculations
do not pose a problem, and although in the present work a small-core
pseudopotential was used for Ag, an all-electron description would also
have been possible. The usage of a pseudopotential has the advantage that 
scalar-relativistic corrections can be included.

\section{Results and Discussion}
\label{potassiumadsorptionsection}

\subsection{Ag(111)   
$(\protect\sqrt{3} \times \protect\sqrt{3})$R30$^\circ$-K}

In this section, the results of the calculations on the system
Ag(111) $(\sqrt{3} \times \sqrt{3})$R30$^\circ$-K
are presented and discussed. 

In table \ref{KonAgroot3},
the adsorption geometries and energies are displayed. 
It turns out that all considered adsorption sites
are virtually degenerate --- the best calculations indicate that the 
binding energies for the various sites are within 0.1 $mE_h$ which is
at the limit of the numerical noise. 
These findings are thus no
contradiction to the experiment where the hcp hollow site was 
identified as the preferential site. The degeneracy is certainly within
the range of the errors associated within the calculation
(lack of basis functions with higher angular momentum than $l=2$, 
number  of Ag layers considered in the slab model, 
uncertainties in the functional, 
anisotropic vibrations observed in the experiment). 
The total computed binding energy per K atom
is 0.042 $E_h$, i.e.  1.1 eV. 

The structure with a stacking fault is also degenerate, and thus
cannot be ruled out by the simulations. In agreement with a recent
calculation \cite{Lietal2002}, the clean Ag(111) surface with a
stacking fault was found to be slightly higher 
in energy (by about 0.00015 $E_h$ per atom) than the
unfaulted surface.

When considering the optimal geometry for the hcp site, 
we note that the computed
distances are in excellent agreement with the experiment: the distance between
the potassium layer and the top silver layer $d_{K-Ag1}$
was found to be 2.79 \AA
\ versus 2.84 $\pm$ 0.03 \AA \ in the experiment \cite{Leathermanetal} and
the distance between first and second silver layer $d_{Ag1-Ag2}$
was obtained to be
2.37 \AA \ (exp.: 2.35 $\pm$ 0.02 \AA). This computed distance
is identical to the distance
in Ag bulk ($4.10/\sqrt{3}$ \AA, with the PBE functional 
\cite{DollHarrison2001}), i.e. no relaxation of the
top layer was found. The distance between potassium and the nearest
silver atom was found to be 3.27 \AA \ (exp.: 3.29 $\pm$ 0.02 \AA). 
We can thus deduce an effective
potassium radius of $3.27 -\frac{4.10}{\sqrt{8}}$\AA=1.82 \AA, in 
 agreement with the experimental value\cite{DiehlMcGrathlang}  
of 1.82 $\pm$ 0.03 \AA. 

The most complex structure included the possibility of
surface rumpling and of lateral relaxation of the top Ag layer.
Rumpling is allowed by symmetry for the bridge and top adsorption site,
and in both cases a large value $\delta 1$ for the rumpling  is obtained
(0.10 \AA \ for the bridge site and 0.22 \AA \ for the top site). The effect
of the rumpling is to push the Ag atom under the adatom deeper into the
bulk, whereas the atom(s) not under the K atom move out of the bulk towards
the K layer.

The importance of rumpling in the second Ag layer under the K layer was
investigated for the hcp hollow site, where this is allowed by symmetry. 
An optimal displacement of
0.01 \AA \ of the Ag atom vertically under the K atom,
away from the K atom, was found, with an energy gain of 0.1 $mE_h$.
In the experiment, a displacement of 0.02 $\pm$ 0.02 \AA \ towards
the K adsorbate layer was found for the Ag atom in the second layer under
the K atom. Although the computed result 
deviates thus slightly from experiment,
 it is important that the
energy gain is fairly small so that second layer rumpling does not appear
to have a huge influence on the results (this was suggested in reference
\onlinecite{Leathermanetal} as a possible reason for the switch in
adsorption site).

Lateral displacement in the top Ag layer plays only a minor role,
and lowers the energy at most by 0.4 $mE_h$. The atoms nearest to the
K adsorbate were allowed to move and the lateral
movement of the atoms was in all cases away from the adsorbate, by
a maximum of 0.04 \AA.

We also note by comparing the results for 4 layers with and without 
surface rumpling, 
that the rumpling is most important for the top site: the
energy gain by rumpling is 2.5 $mE_h$, and 0.5 $mE_h$ for the bridge site.
This is in line with previous calculations for K/Cu(111) \cite{Doll2001KCu}.

\subsection{Ag(111)   $(2 \times 2)$-K}

In the following table 
\ref{KonAg2x2}, results for the Ag(111)$(2 \times 2)$-K
structure are displayed. 
Fcc hollow, hcp hollow and bridge site
are within 0.2 $mE_h$, i.e. practically degenerate. The structure with
a stacking fault is $\sim$ 0.8 $mE_h$ higher in energy, the top site
by $\sim$ 1.8 $mE_h$. Similar as for the higher coverage, it can only
be stated that these findings are compatible with the experiment, but
a statement about the preferred site is not possible.

The geometry is in good agreement with experiment:
a value for the  interlayer distance $d_{K-Ag1}$ of
 2.60 \AA \ (exp.: 2.70 $\pm$ 0.03 \AA)  was obtained
and an interlayer distance $d_{Ag1-Ag2}$ 
between the first and second silver layer
of  2.35 \AA \ (exp.: 2.34 $\pm$ 0.02 \AA). Surface rumpling is
important and a value of $\delta 1=$0.11 \AA \
was computed (exp.: 0.10 $\pm$ 0.03 \AA).
The effect of the surface rumpling was again to push those atom(s)
under the potassium
adsorbate deeper into the substrate, whereas the other atom (three atoms
in the case
of the top site, two atoms in the case of the bridge site and one atom
 in the case of
the threefold hollow sites) move outwards towards the potassium layer.

Lateral relaxation plays a minor role for the binding energy. In all cases
except for the top site where this is not possible,
the atoms nearest to K move away from the K atom; in the case of the
top site, the three atoms not under the K atom move slightly
towards each other,
away from the K atom. However, the energy gain is at the limit of the
numerical accuracy.

For comparison, calculations without surface rumpling were performed,
i.e. $\delta 1$ was kept fixed at 0 \AA.
We note that the interlayer distances 
$d_{K-Ag1}$ between
K layer and top Ag layer increase when rumpling is not possible, and are
2.73 \AA \ for fcc and hcp site. 
The interlayer distance between
first and second Ag layer hardly changes. The adsorption energy differs
by  $\le 1 mE_h$ for fcc, hcp, bridge site and for the structure with
a stacking fault; but by 2.9 $mE_h$
for the top site. Thus we find again that surface rumpling is most
important for the top site, however, it is not big enough to make
the top site the preferred adsorption site.

The computed adsorption energy per potassium atom 
for this structure 
is 0.041 $E_h$, i.e. nearly identical to the higher coverage.
This energy is close to the one computed for K/Al(111); also the 
splitting between various sites was found to be of the order of few 
hundreths of an eV only \cite{NeugebauerScheffler}; the 
binding energy is also
comparable to Cu(111)$(2 \times 2)$-K \cite{Doll2001KCu}.
The adsorption
energy is however lower than the one computed for K/Pt(111) where
values of 2.42 eV (at a coverage of
one third of a monolayer) and 2.93 eV (at a coverage of
one fourth of a monolayer) were obtained, at the level of the local
density approximation\cite{Moreetal}. The energy splitting between fcc
and hcp site was computed to be 15 meV in the latter publication, with
the hcp site being lower; this is compatible with the near-degeneracy
which is found here for the system K/Ag(111).

Finally, we also see that the results are well converged with respect to
the number of layers, for both coverages: 
in one case (substrate rumpling, no lateral relaxation), the optimization
has been performed with 4 and 5 Ag layers, 
and both geometry and adsorption energies are stable.

\subsection{Population and density of states}

In table \ref{Kpopulationtable}, Mulliken charges for the potassium
atom are given, projected on the different basis functions. 
These charges are defined as

\[ N_{A}=\sum_{\mu \in A}\sum_{\nu}P_{\mu\nu}S_{\nu\mu} \]

with density matrix $P$, overlap matrix $S$ and $A$ the set of basis 
functions for which the population is computed.

It should
be mentioned that a relatively large number of digits has been given;
although the absolute values will not be more accurate than to one or two
decimal points, for the relative values more digits are still meaningful.

In the case of Ag(111) $(\sqrt{3} \times \sqrt{3})$R30$^\circ$-K,
the total charge is $\sim$ 0.16 $|e|$, in the case of 
Ag(111) $(2 \times 2)$-K $\sim$ 0.24 $|e|$. Obviously, if the
charges were the same, the
electrostatic repulsion at higher coverage 
(Ag(111) $(\sqrt{3} \times \sqrt{3})$R30$^\circ$-K)
would be larger because of
the shorter K-K distances. Therefore, the charge in the case of
Ag(111) $(\sqrt{3} \times \sqrt{3})$R30$^\circ$-K is reduced, in
agreement with the Langmuir-Gurney model.
The individual populations are very similar to Cu(111) $(2 \times 2)$-K:
$p_x$, $p_y$ and $p_z$ charges are slightly above 4, the number of 
$s$-electrons is $\sim$ 6.5;
charges in $p_x$ and $p_y$ orbitals are identical except for the bridge
site;
the $p_x$ orbital which has more overlap
with the substrate layer (in our choice of geometry)
is slightly less occupied than the $p_y$ orbital. 
The higher occupancy of that
orbital, which has less overlap with the substrate, indicates that
the bond is not covalent. In addition,
the overlap population defined as 

$\sum_{\mu \in A}\sum_{\nu \in B}P_{\mu\nu}S_{\nu\mu}$, 

with $A$ the set of
basis functions on the first atom and $B$ the set of basis functions of the
other atom, was computed. This number is
a measure for covalency, and 
is, similar to Cu(111)(2$\times$2)-K, very small: for K and nearest
neighbor Ag, it is $\le$ 0.06 $|e|$ for both coverages, 
and negligible for further neighbors. 

\begin{figure}
\caption{DOS, projected on K, fcc adsorption site, 
$(\protect\sqrt{3}\times\protect\sqrt{3})$R30$^\circ$ structure. 
The DOS, projected on all K basis functions, 
is shown, together with the DOS projected on $s$, $p_x$, $p_y$ and $p_z$ 
orbitals only.
The Fermi energy
is indicated with a vertical line.}
\vspace{1cm}
T=0.005
\label{DOSfigureroot3}
{\psfig
{figure=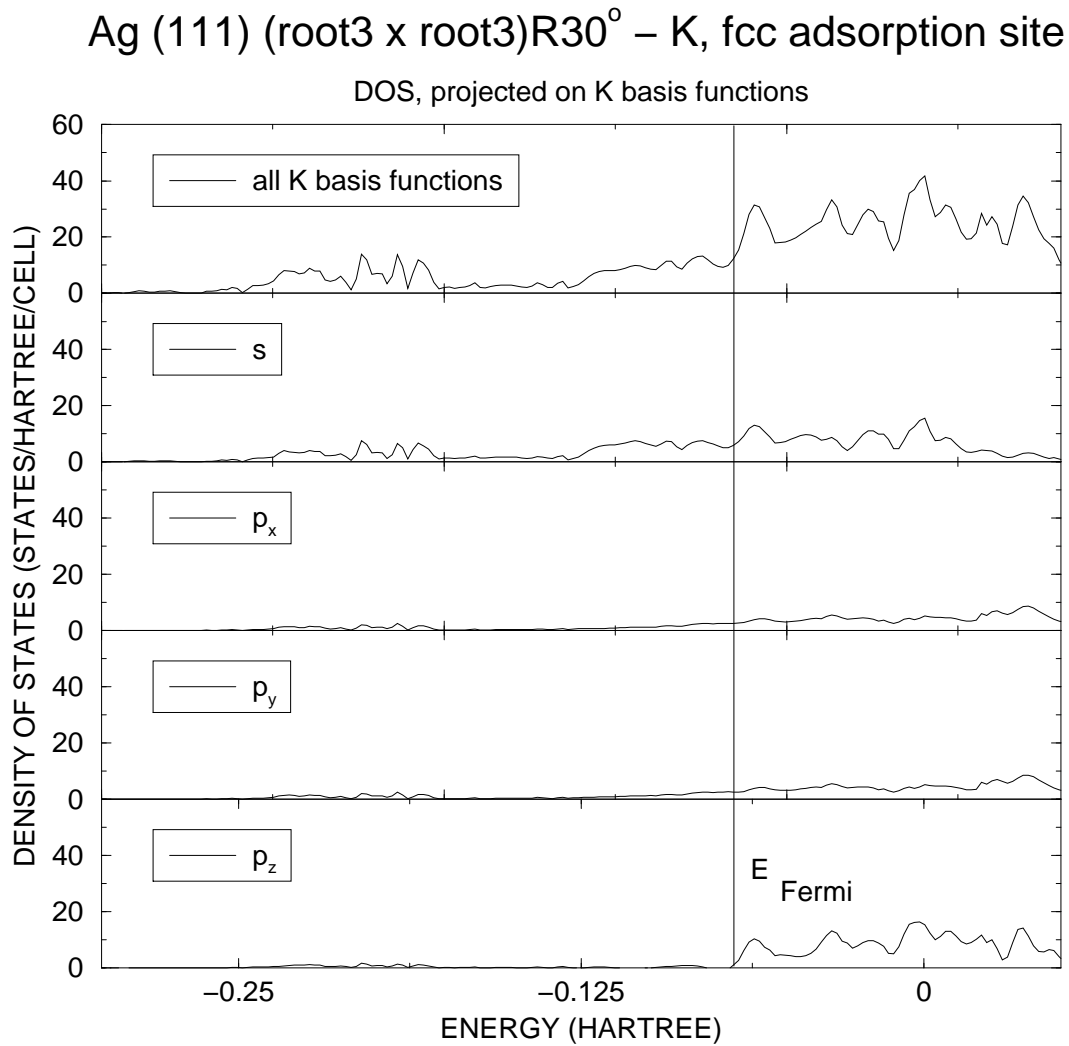,width=15cm,angle=270}}
\end{figure}

\begin{figure}
\caption{DOS, projected on K, fcc adsorption site, 
(2x2) structure. 
The DOS, projected on all K basis functions, 
is shown, together with the DOS projected on $s$, $p_x$, $p_y$ and $p_z$ 
orbitals only.
The Fermi energy
is indicated with a vertical line.}
\vspace{1cm}
\label{DOSfigure2x2}
{\psfig
{figure=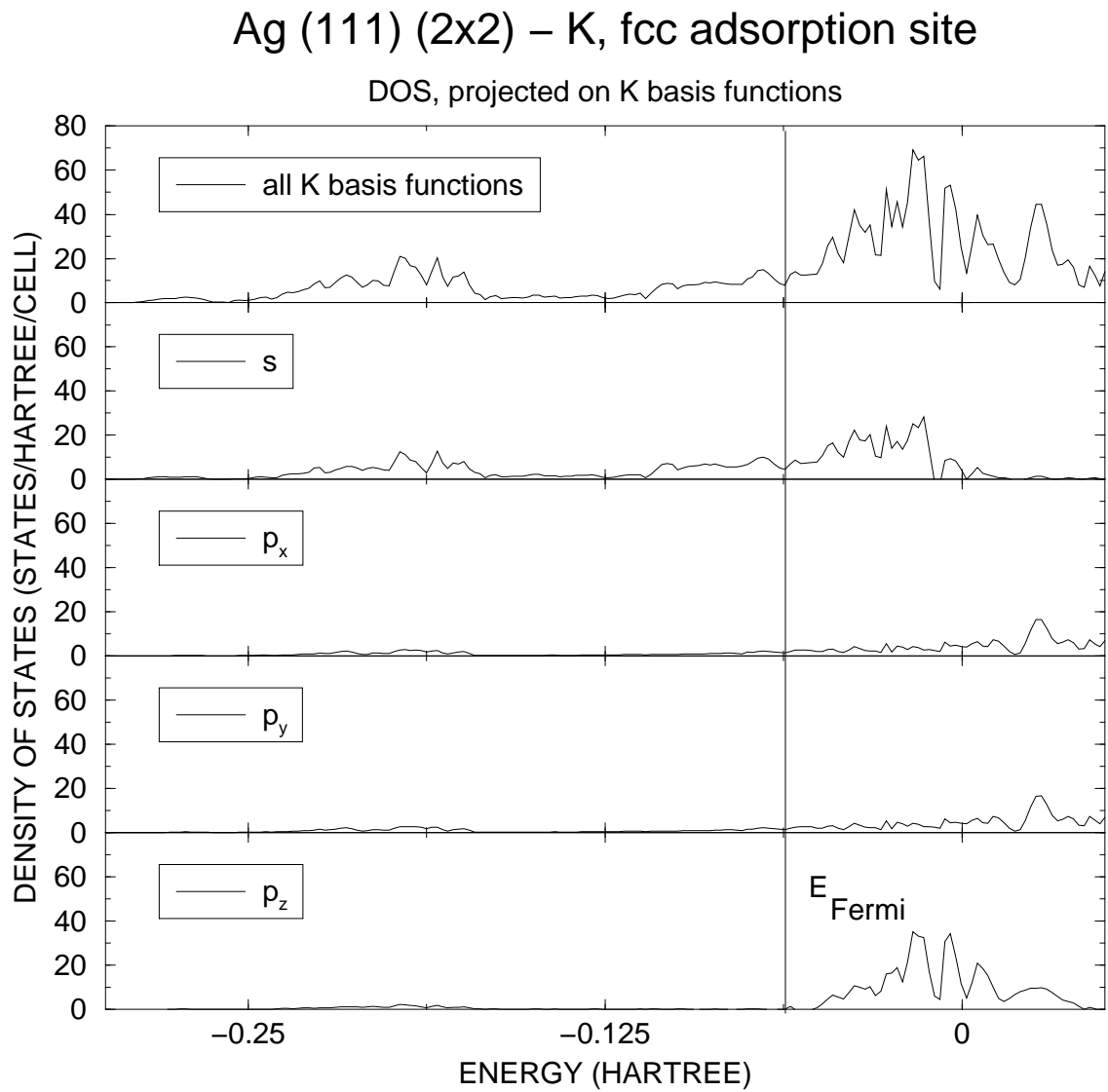,width=15cm,angle=270}}
\end{figure}

In figures \ref{DOSfigureroot3} and \ref{DOSfigure2x2}, 
the density of states (DOS), projected on the
potassium basis functions, is displayed for the fcc adsorption site. 
The Fermi
energy is in a regime with a significant contribution  from K 4$s$ and
4$p$ bands, so that the overlayer is clearly metallic. 
The projected DOS does not depend
on the adsorption site: it looks virtually
identical for fcc hollow, hcp hollow, bridge and top site, and also for
the structure with stacking fault.
This is as found for Cu(111)(2 $\times$ 2)-K, but 
different from the case of chlorine as an adsorbate, where 
the projected DOS clearly depended on the
adsorption site\cite{DollHarrison2001}. 

The values
of the K $3s$
and $3p$ levels 
are practically independent of the adsorption
site (see table \ref{coreleveltable}).
This is again in contrast to chlorine: for chlorine, the core eigenvalue
depended on the adsorption site, and it was correlated with the charge of
the chlorine atom \cite{DollHarrison2001}. For K, the variation of
the  Mulliken charges with adsorption site for fixed coverage
is much smaller (table
\ref{Kpopulationtable}) as for chlorine. This is thus consistent with
the finding that the core eigenvalues do not vary for the different
sites. Also, the position of the Fermi energy and thus the value of
the work function is independent of the site, and changes slightly with
the coverage.

\subsection{Comparison of the results for the two coverages}

When comparing the results for the two coverages, we firstly note that
 the bond lengths
for the individual sites are very similar for both coverages. The 
bond length increases with the number of nearest neighbors to which the
potassium adatom bonds, in agreement with Pauling's argument \cite{Pauling}
that a smaller number of bonds leads to a stronger individual bond and a
shorter bond length: the bond length
is shortest for the top site with 2.95 \AA \ for the
$(\sqrt{3} \times \sqrt{3})$ structure
(or 2.89 \AA \ for the $(2 \times 2)$ structure), 
longer for the bridge site
and longest for the threefold hollow sites. Similarly, the effective K radius,
obtained from the difference of bond length and effective silver radius
($\frac{4.10}{\sqrt{8}}$\AA ),
increases in this order. If we keep the Ag radius fixed, 
it increases from 1.50 \AA \ (or 1.44 \AA 
\ for the $(2 \times 2)$ structure) for the top site
 to 1.82 \AA \ (or 1.76 \AA)
for the threefold hollow sites. These values are compatible with
effective K radii deduced from experiment (see table 10 in reference
\onlinecite{DiehlMcGrathlang}). 

In table \ref{Vergleichstabelle}, the most important
results of the calculations
for the two coverages and the fcc hollow are compared. 

The binding energy per K atom, with respect to free K atoms and
a clean Ag(111) slab,  is 0.042 and 0.041 $E_h$, i.e. nearly identical
 for both coverages. In the case of
Ag (111) $(\sqrt{3} \times \sqrt{3})$R30$^\circ$-K, all considered sites,
and in the case of Ag(111) $(2 \times 2)$-K, the fcc, hcp and bridge
site are virtually degenerate. This degeneracy was already observed
in the early study of K/Al(111) \cite{NeugebauerScheffler}
and explained with the fact that the large radius of potassium leads
to a large distance of the adsorbate to the substrate and
therefore the K adsorbate will experience only
a small substrate electron density corrugation; similarly the potassium
overlayer is metallic and its electronic charge more diffuse
than the charge of, e.g. chlorides \cite{DollHarrison2000,DollHarrison2001}
 or oxides \cite{Lietal2002}
as adsorbates which also helps to
reduce the energy difference between the threefold hollow sites 
and bridge or top site, compared to halogenides or oxides.

The K charge hardly depends on the adsorption site, but it depends
on the coverage. The slightly higher charge (i.e. less electrons) for the
lower coverage is consistent with the slightly shorter bond
length and smaller
effective K radius for this adsorbate system. 

The values of the 
K $3s$ and $3p$ core eigenvalues also indicate that there is a little
difference between the two coverages: For the lower coverage, the
core eigenvalues are at 1.194 $E_h$ and 0.590 $E_h$ below the Fermi level,
for the higher coverage at 1.184 $E_h$ and 0.600 $E_h$. These values
are virtually independent of the adsorption site. Again, this is consistent
with the finding that the K charge is larger for the lower coverage
and therefore the core eigenvalues are lower (i.e. there is slightly
less electronic
charge to screen the nuclear charge and the core levels
are thus further stabilized).

The work function decreases from 0.131 $E_h$ for the clean Ag(111) surface
to 0.062 $E_h$ for the (2 $\times$ 2) structure and increases again slightly
to 0.069 $E_h$ for the $(\sqrt{3} \times \sqrt{3})R30$ structure.
This finding is in line with the argument that initially, the work
function will decrease linearly with increasing coverage, 
up to a minimum, and finally increase again
because of the depolarization of the adsorbate (see, e.g. ref. 
\onlinecite{Zangwill}).

Comparing the results for K/Ag(111) with K/Cu(111) and the (2 $\times$
2) pattern, we note that
the site has switched from top to the threefold hollow or bridge site.
The reason appears to be that, without rumpling, the energy splitting
between hollow or bridge sites and top site is less than 1 $mE_h$ for 
K/Cu(111) \cite{Doll2001KCu}, 
but 4 $mE_h$ for K/Ag(111). Substrate rumpling changes this
splitting in favor of the top site; however, the change is not large enough
to stabilize the top site as the ground state for K/Ag(111). 
The parameter most important 
for the magnitude of the energy splitting of the sites, without 
rumpling, appears to be the nearest neighbor spacing of the substrate: 
for a fixed radius of the
K adsorbate, the adsorbed atom will have more overlap with more substrate
atoms, if the substrate atoms are closer. In the case of the top site,
the K atom has a nearest neighbor vertically below and six next nearest
neighbors in the first layer. These six atoms are closer in the case
of K/Cu(111) than K/Ag(111) because of the smaller atomic radius of
copper and thus smaller lattice constant of Cu. Thus, the energy splitting
is smaller for K/Cu(111) compared to K/Ag(111), 
and substrate rumpling can change it in favor of the top
site for K/Cu(111). This argument would also hold for the
 system K/Ru(0001) \cite{Gierer1992} which behaves similar to K/Ag(111).
The calculations
thus support the argument given, for example, in reference
\onlinecite{DiehlMcGrathlang} and which was based on
a large set of experimental data 
(when varying the adsorbate, the ratio of the radius of the adsorbate
to the nearest neighbor spacing of the substrate was recommended as a
parameter to predict the adsorption site). 
In the simulations performed here,
this argument can be verified by comparing the results from
 a surface without substrate rumpling
and a surface with rumpling.

For substrates with
similar lattice constants, an additional parameter will be the
energy required for the deformation of the surface; this energy is
found to be 1.5 $mE_h$ for K/Cu(111) and 2.8 $mE_h$ for K/Ag(111), for
the top site ((2 $\times$ 2) structure, the energy difference between the
unrelaxed clean surface and the optimal geometry of the adsorbate system, 
without K adsorbate, was computed). Note that the magnitude of
the rumpling is larger for K/Ag(111), which may explain the higher
energy necessary for the deformation of the clean Ag surface
(although Ag has a lower bulk modulus than Cu). 

We also note that the top site is slightly better in energy for the
higher coverage than for the lower coverage, 
relative to the other sites (in line with the case of 
K/Al(111) \cite{NeugebauerScheffler}).
This could be due to
the larger radius of the K adsorbate for the higher coverage. As explained
earlier, the positive K charge is smaller for the higher coverage and thus the
radius is larger. Therefore, the overlap with the 6 nearest neighbors
will be larger and the top site will be slightly more stabilized.
However, this argument must be contrasted with the case of Cs/Ru(0001)
\cite{OverPRB1992} where the site switches from top 
((2 $\times$ 2) structure) to hcp 
($(\sqrt{3} \times \sqrt{3})$R30$^\circ$ structure). In the latter
case, a better screening was suggested as the possible explanation.
These two controversial findings for similar systems
demonstrate how difficult it is to give even only 
qualitative rules for the preferred adsorption site.

We can thus give a reasonable argument for the question, why the top site is
not preferred for K/Ag(111), in contrast to K/Cu(111). 
The question of the site switch from fcc
to hcp, however, can not yet be answered. 
The difficulty results from the small energy splitting, and thus large
number of effects on this energy scale that are possible, and may
influence the site preference (e.g. slightly different charge,
screening, second layer rumpling, stacking fault). Numerically, the
energy splitting is at the limit of the numerical noise.

\section{Summary}

In summary, the impact of the article is twofold: firstly, the
technology of using Gaussian type basis functions was shown to be 
capable of describing the adsorption on metallic surfaces. Secondly,
some of the experimental findings could be confirmed (essentially
the geometry), and new data could be provided (energy splitting between
the various adsorption sites, Mulliken charges, 
core eigenvalues, work functions, and the dependence of these parameters
on the coverage).

In more detail, 
it has been shown that density functional calculations can reproduce
a part of the experimental findings for the systems 
Ag (111) $(\sqrt{3} \times \sqrt{3})$R30$^\circ$-K
and Ag(111) (2$\times$2)-K. The computed
geometry is in excellent agreement with
the experimental results, and therefore much better than for
the case of K/Cu(111) where there was a deviation of $\sim$ 0.2  \AA \ 
between the experiment and the simulation \cite{Adler1993,Doll2001KCu}.
This agreement for K/Ag(111) thus helps to support the validity of
the approach employed in the simulations. 

The adsorption site was found to be nearly
degenerate for various sites considered for both coverages, and a value of
0.04 $E_h$ was obtained for the binding energy. 
The main difference to the system K/Cu(111), where the top site is occupied,
appears to be due the different lattice constant of Cu vs. Ag. In both cases,
a threefold hollow site would be favored without substrate rumpling,
but only in the case of K/Cu(111), the energy gain associated with
the rumpling is large enough to lead to a change of the adsorption site and
makes the top site favorable.

The computed energies demonstrate that various sites are virtually 
degenerate, within the accuracy of the calculations. This appears 
to be due to the metallic nature of the overlayer and the large
radius of the potassium adsorbate, 
which makes the surface electron-density
corrugation have only little impact on the energy splitting
between the various sites. Possible 
suggestions for the change in adsorption site from fcc to hcp 
were tested (second layer rumpling, stacking fault). However, second layer
rumpling did
not have a huge impact, and the energy of the structure
with a stacking fault was still degenerate with the other structures, within
the accuracy of the calculations. 
It is therefore possible to give arguments why the top site is not occupied,
but the question of the site change from fcc to hcp remains puzzling.
The near-degeneracy of the energies indicates that minute effects are
responsible for the site change.

The computed potassium charge is between +0.16 $|e|$ for the higher
and +0.24 $|e|$ for the lower coverage. These charges are
consistent with results for the effective K radius and K-Ag bond length
which increase for the higher 
coverage, with the work function, and with
the K $3s$ and $3p$ core eigenvalues
which move slightly closer to the Fermi energy for the higher coverage.
This consistency indicates that Mulliken charges are a simple and
useful concept for describing adsorption on metallic surfaces.
As in the case of K/Cu(111), the small overlap population indicates that
virtually no evidence can be found for a covalent contribution to the
binding.


\appendix*
\section{Numerical Integration}

The accuracy of the numerical integration depends especially on the number of
sampling points and the smearing temperature employed. Obviously, a higher
number of sampling points will lead to a higher accuracy. The optimal smearing
temperature will depend on the band structure and on the number of 
$\vec k$-points. In general, a lower smearing temperature will be better
because the Fermi function (which is a step 
function at zero temperature and thus makes accurate 
integration numerically difficult) is better approximated.
However, when quantities such as susceptibilities
are computed, the smearing temperature should not be too low, because
the function to be integrated has singularities (see, e.g., the discussion in
reference \onlinecite{Gillan}). 

To investigate the dependence of the results on the smearing temperature,
first a simple test was performed where the lattice constant of bulk
Ag was optimized at $T=0.005 E_h$. The optimal lattice constant was
4.12 \AA, the cohesive energy 0.090 $E_h$ and the bulk modulus 108 GPa.
At a higher temperature of $T=0.03 E_h$, the optimized values were
\cite{DollHarrison2001}
4.10 \AA, 0.088 $E_h$ and 113 GPa. The  higher smearing temperature 
of 0.03 $E_h$ thus does not seem to have a huge impact on the result.
This is also seen in the comparison in tables \ref{KonAgroot3} and
\ref{KonAg2x2}, where the energy splitting between the various sites
hardly changes when the smearing temperature is reduced from 0.03 $E_h$
to 0.005 $E_h$. There is a change in the total binding energy, which is
however still negligible compared to its magnitude.

\begin{figure}
\caption{Binding energy 
$E_{\rm {K/Ag(111)}}-{E_{\rm Ag(111)}-E_{\rm K}}$,
for K adsorbed on the hcp site of the (111) surface of a 3 layer Ag slab.
Various smearing temperatures from 0.001 $E_h$ to 0.05 $E_h$, and 
$\vec k$-point samplings from 2$\times$2 up to 16$\times$16 have been 
compared.}
\vspace{1cm}
\label{KAghcpKTfigure}
{\psfig
{figure=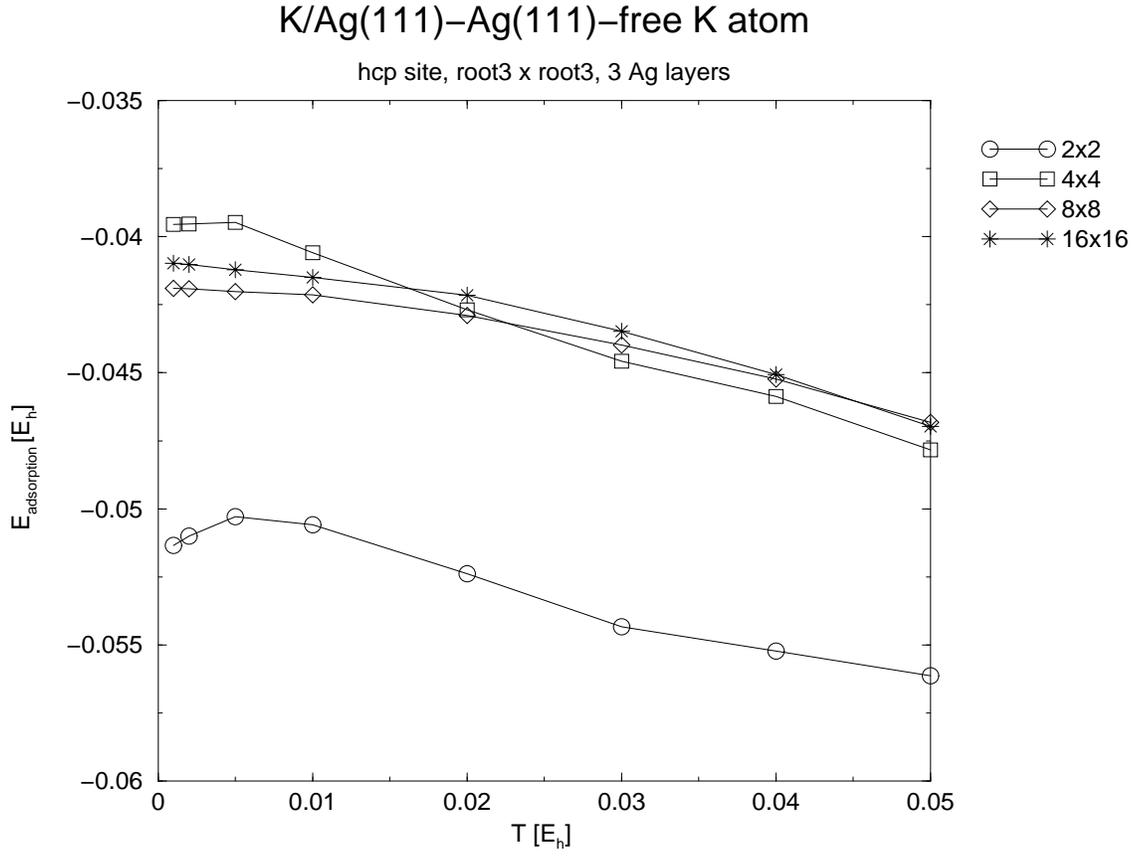,width=15cm,angle=270}}
\end{figure}

Finally, in figure \ref{KAghcpKTfigure}, the binding energy 
$E_{K/Ag(111)}-E_K-E_{Ag(111)}$
is computed for the hcp site, with a 3 layer slab simulating the Ag(111)
surface. It becomes obvious that the binding energy depends similarly
on the number of $\vec k$-points as on the smearing temperature. 
As a whole, it can be concluded that the employed smearing temperature
should not lead to serious errors for the results presented in this
article.

\onecolumn

\newpage
\begin{table}
\begin{center}
\caption{Adsorption of K on the Ag(111) surface,
$({\protect\sqrt{3}}  \times \protect\sqrt{3} )$ R30$^\circ$ pattern.
The adsorption energy $E_{ads}$ is the difference 
$E_{\rm {K/Ag(111)}}-{E_{\rm Ag(111)}-E_{\rm K}}$.
For the most complex geometry, this energy was 
computed with two different parameters of the smearing temperature,
0.03 $E_h$ and 0.005 $E_h$, for comparison.}
\label{KonAgroot3} 
\begin{tabular}{cccccccc}
site & $Bond \ length$ & 
$d_{K-Ag1}$ & $d_{Ag1-Ag2}$ & $\delta 1 $ & lateral &
$E_{ads}(0.03)$ & $E_{ads} (0.005)$  \\
 & [\AA] &  [\AA] & [\AA]  & [\AA]  & [\AA]  
& $\left[\frac{E_h}{K \ atom}\right]$ & $\left[\frac{E_h}{K \ atom}\right]$ \\
\\
\multicolumn{8}{c}{4 layers, without substrate rumpling, without lateral relaxation}\\
fcc hollow & 3.26  & 2.80 & 2.37 & 0 & 0 & -0.0442 \\
hcp hollow & 3.26  & 2.80 & 2.37 & 0 & 0 &  -0.0440\\
bridge &     3.16  & 2.81  & 2.37 & 0    & 0 &  -0.0437 \\
top &        2.94  & 2.94 &   2.37   & 0 & 0 &  -0.0413\\
stacking fault & 3.27  & 2.81 & 2.38 & 0 & 0 & -0.0435 \\
\\
\multicolumn{8}{c}{4 layers, with substrate rumpling, without lateral relaxation} \\
fcc hollow & 3.26  & 2.80 & 2.37 & 0 & 0 &  -0.0442 \\
hcp hollow & 3.26  & 2.80 & 2.37 & 0 & 0 &  -0.0440\\
bridge &     3.18  & 2.73  & 2.34 & 0.10    & 0 &  -0.0442 \\
top &        2.96  & 2.73 &   2.24   & 0.23 & 0 &  -0.0438 \\
stacking fault & 3.27  & 2.81 & 2.38 & 0 & 0 & -0.0435 \\
\\
\multicolumn{8}{c}{5 Ag layers, with substrate rumpling, without lateral relaxation} \\
fcc hollow & 3.27  & 2.81 & 2.37 & 0    & 0 & -0.0443 \\
hcp hollow & 3.27  & 2.81 & 2.37 & 0    & 0 & -0.0441 \\
bridge &     3.18  & 2.74 & 2.34 & 0.10 & 0 & -0.0443 \\
top &        2.95  & 2.73 & 2.25 & 0.22 & 0 & -0.0441\\
stacking fault & 3.27  & 2.81 & 2.38 & 0 & 0 & -0.0437 \\
\\
\multicolumn{8}{c}{5 Ag layers, with substrate rumpling, with lateral relaxation} \\
fcc hollow & 3.27  & 2.79 & 2.37 & 0 & 0.04  & -0.0444 & -0.0422 \\
hcp hollow & 3.27  & 2.79 & 2.37 & 0 & 0.04  & -0.0442 & -0.0421\\
bridge &     3.19  & 2.74 & 2.33 & 0.10 & 0.01 &-0.0444 & -0.0422\\
top &        2.95  & 2.73 & 2.25 & 0.22 & 0 & -0.0441 & -0.0422 \\
stacking fault & 3.27  & 2.79 & 2.38 & 0 & 0.03 & -0.0441 & -0.0422 \\
\\
exp (hcp) \cite{Leathermanetal,Kaukasoinaetal} 
          & 3.29 $\pm$ 0.02 & 2.84 $\pm$ 0.03 & 2.35 $\pm$ 0.02 & 0\\
\end{tabular}
\end{center}
\end{table}

\begin{table}
\begin{center}
\caption{Adsorption of K on the Ag(111) surface, $(2  \times 2 )$  pattern.
The adsorption energy  $E_{ads}$ is the difference 
$E_{\rm {K/Ag(111)}}-{E_{\rm Ag(111)}-E_{\rm K}}$.
For the most complex geometry, this energy was 
computed with two different parameters of the smearing temperature,
0.03 $E_h$ and 0.005 $E_h$, for comparison. }
\label{KonAg2x2} 
\begin{tabular}{cccccccc}
site & $Bond \ length$ & $d_{K-Ag1}$  & $d_{Ag1-Ag2}$ & $\delta 1 $
& lateral & $E_{ads}(0.03)$ & $E_{ads}(0.005)$  \\
& [\AA] & [\AA] & [\AA] &  [\AA] & [\AA] & 
$\left[\frac{E_h}{K \ atom}\right]$ & $\left[\frac{E_h}{K \ atom}\right]$ \\
\\
\multicolumn{8}{c}{4 Ag layers, without substrate rumpling, without lateral
relaxation} \\
fcc hollow &   3.20 &  2.73 & 2.37 & 0 & 0 & -0.0418 \\
hcp hollow &   3.20 &  2.73 & 2.37 & 0 & 0 & -0.0416\\
bridge     &   3.11 &  2.75 & 2.37 &  0 & 0 &-0.0412\\
top        &   2.87 &  2.87 & 2.36 &  0 & 0 &-0.0378\\
stacking fault &   3.21 & 2.74  & 2.39 & 0 & 0 & -0.0409 \\
\\
\multicolumn{8}{c}{4 Ag layers, with substrate rumpling, without lateral
relaxation} \\
fcc hollow & 3.21  &  2.61 & 2.34 & 0.13 &  0 & -0.0424 \\
hcp hollow & 3.21  &  2.61 & 2.34 & 0.13 &  0 & -0.0422 \\
bridge &     3.12  &  2.64 & 2.31 & 0.12 &  0 & -0.0422 \\
top &        2.90  &  2.67 & 2.22 & 0.23 &  0 & -0.0407 \\
stacking fault & 3.21  & 2.61 &  2.35 & 0.13 & 0 & -0.0415 \\
\\
\multicolumn{8}{c}{5 Ag layers, with substrate rumpling, without lateral
relaxation} \\
fcc hollow & 3.20  &  2.61 & 2.34 & 0.12 &  0 & -0.0423 \\
hcp hollow & 3.21  &  2.61 & 2.34 & 0.13 &  0 &  -0.0423 \\
bridge &     3.12  &  2.66 & 2.32 & 0.11 &  0 &  -0.0422 \\
top &        2.89  &  2.66 & 2.22 & 0.23 &  0 &  -0.0408 \\
stacking fault & 3.21  &  2.62 & 2.35 & 0.12 &  0 &  -0.0415 \\ 
\\
\multicolumn{8}{c}{5 Ag layers, with substrate rumpling, with lateral relaxation}
 \\
fcc hollow & 3.20  &  2.60 & 2.35 & 0.11 &    0.04 & -0.0424 & -0.0411 \\
hcp hollow & 3.21  &  2.60 & 2.35 & 0.12 &    0.03 &  -0.0425 & -0.0413\\
bridge &     3.13  &  2.65 & 2.32 & 0.11 &    0.04  & -0.0425 & -0.0411 \\
top &        2.89  &  2.66 & 2.22 & 0.23 &    0.005 & -0.0408 & -0.0395 \\
stacking fault & 3.20  &  2.59 & 2.35 & 0.13 & 0.02 & -0.0417 & -0.0405
\\
exp (fcc) \cite{Leathermanetal,Kaukasoinaetal} 
  & 3.27 $\pm$ 0.03 & 2.70 $\pm$ 0.03 & 2.34 $\pm$ 0.02 & 0.10 $\pm$ 0.03 \\
\end{tabular}
\end{center}
\end{table}

\begin{table}
\begin{center}
\caption{Orbital-projected potassium charge for different adsorption sites
(5 Ag layers, with substrate rumpling and lateral relaxation).}
\label{Kpopulationtable}
\begin{tabular}{ccccccc}
site & \multicolumn{6}{c}{charge, in $|e|$} \\
& $s$ & $p_x$ & $p_y$ & $p_z$ & total \\
 & \multicolumn{6}{c}{$(\sqrt{3}\times\sqrt{3})$R30$^\circ$}\\
fcc hollow   & 6.513 & 4.135 & 4.135 & 4.052  & 18.843 \\
hcp hollow   & 6.515 & 4.134 & 4.134 & 4.052  & 18.843 \\
bridge       & 6.513 & 4.127 & 4.144 & 4.052  & 18.845 \\
top          & 6.502 & 4.146 & 4.148 & 4.051  & 18.856 \\
stacking fault   & 6.515 & 4.134 & 4.134 & 4.051  & 18.843 \\
 & \multicolumn{6}{c}{(2 $\times$ 2)}\\
fcc hollow   & 6.494 & 4.115 & 4.114 & 4.025  & 18.757 \\
hcp hollow   & 6.496 & 4.114 & 4.114 & 4.025  & 18.757 \\
bridge       & 6.496 & 4.106 & 4.122 & 4.027  & 18.760 \\
top          & 6.487 & 4.123 & 4.123 & 4.028  & 18.772 \\
stacking fault   & 6.496 & 4.114 & 4.114 & 4.025  & 18.757 \\
\end{tabular}
\end{center}
\end{table}

\begin{table}
\begin{center}
\caption{Position of the K $3s$ and $3p $ core eigenvalues, and position of
the Fermi energy, in $E_h$ 
(5 Ag layers, with substrate rumpling and lateral relaxation).}
\label{coreleveltable}
\begin{tabular}{ccccccc}
site & $(\sqrt{3}\times\sqrt{3})$R30$^\circ$ & (2 $\times$ 2)\\
fcc hollow   & -1.253 ; -0.659 ; -0.069 & -1.256 ; -0.662 ; -0.062\\ 
hcp hollow   & -1.253 ; -0.658 ; -0.069 & -1.256 ; -0.662 ; -0.062 \\ 
bridge       & -1.253 ; -0.659 ; -0.069 & -1.257 ; -0.662 ; -0.062 \\ 
top          & -1.253 ; -0.659 ; -0.069 & -1.258 ; -0.664 ; -0.062 \\ 
stacking fault   & -1.253 ; -0.659 ; -0.069 & -1.256 ; -0.661 ; -0.062 \\ 
\end{tabular}
\end{center}
\end{table}


\begin{table}
\begin{center}
\caption{Comparison of the results for the two coverages (fcc hollow site).}
\label{Vergleichstabelle}
\begin{tabular}{ccc}
structure & $(\sqrt{3}\times\sqrt{3})$R30$^\circ$-K & $(2 \times 2)$-K \\
bond length [\AA]& 3.27 & 3.20  \\
effective K radius [\AA] & 1.82 & 1.76 \\
binding energy $\left[\frac{E_h}{K \ atom}\right]$ & 0.042 & 0.041\\
Mulliken charge on K $[|e|]$ & 0.16 & 0.24 \\
K $3s$, $3p$ core eigenvalues, relative to Fermi energy
 $[E_h]$ & -1.184 ; -0.590 & -1.194 ; -0.600 \\
work function $[E_h]$ (clean Ag(111): 0.131) & 0.069  & 0.062  \\
\end{tabular}
\end{center}
\end{table}

\newpage

\newpage

\newpage

\newpage

\newpage

\newpage

\end{document}